\documentclass{article}
\usepackage{spconf,amsmath,graphicx,epsfig,epstopdf}

\title{Gaussian-Constrained training for speaker verification}
\name{Lantian Li, Zhiyuan Tang, Ying Shi, Dong Wang\thanks{
This work was supported by the National Natural Science
Foundation of China No. 61633013,
and the Postdoctoral Science Foundation of China No. 2018M640133.
L.L.T and T.Z.Y are joint first authors.
Dong Wang is the corresponding author (wangdong99@mails.tsinghua.edu.cn).}}
\address{Center for Speech and Language Technologies, RIIT, Tsinghua University, China\\
Beijing National Research Center for Information Science and Technology}%
\begin{document}
%
\maketitle
\begin{abstract}
Neural models, in particular the d-vector and x-vector architectures, have produced state-of-the-art performance on many speaker verification tasks.
However, two potential problems of these neural models deserve more investigation.
Firstly, both models suffer from `information leak', which means that some parameters participating in model training will be discarded during inference,
i.e, the layers that are used as the classifier.
Secondly, these models do not regulate the distribution of the derived speaker vectors. This
`unconstrained distribution' may degrade the performance of the subsequent scoring component, e.g., PLDA.
This paper proposes a Gaussian-constrained training approach that (1) discards the parametric classifier, and (2) enforces the distribution of
the derived speaker vectors to be Gaussian. Our experiments on the \emph{VoxCeleb} and \emph{SITW} databases demonstrated
that this new training approach produced more representative and regular speaker embeddings,
leading to consistent performance improvement.

\end{abstract}
\section{Introduction}
\label{sec:intro}

Automatic speaker verification (ASV) is an important biometric authentication technology and
has found a broad range of applications. The current ASV methods can be categorized
into two groups: the statistical model approach that has gained the most
popularity~\cite{Reynolds00,Kenny07,dehak2011front}, and the neural model
approach that emerged recently but has shown great potential~\cite{ehsan14,heigold2016end,li2017deep}.

Perhaps the most famous statistical model is the Gaussian mixture model$-$universal background
model (GMM-UBM)~\cite{Reynolds00}. It factorizes the variance of speech signals
by the UBM, and then models individual speakers conditioned on that factorization.
More succinct models design subspace structures to improve the statistical strength, including
the joint factor analysis model~\cite{Kenny07} and the i-vector model~\cite{dehak2011front}.
Further improvements were obtained by either discriminative models (e.g., PLDA~\cite{Ioffe06}) or phonetic knowledge
transfer (e.g., the DNN-based i-vector model~\cite{Kenny14,lei2014novel}).

The neural model approach has also been studied for many years as well,
however it was not as popular as the statistical model approach until recently
training large-scale neural models became feasible.
The initial success was reported by Ehsan et al. on a text-dependent
task~\cite{ehsan14}, where frame-level speaker features were extracted from the last hidden layer
of a deep neural network (DNN), and utterance-based speaker vectors (`d-vectors') were derived by
averaging the frame-level features. Learning frame-level speaker features
offers many advantages, which paves the way to deeper understanding of speech signals.

Researchers followed Ehsan's work in two directions. In the first direction, more speech-friendly
DNN architectures were designed, with the goal of learning stronger frame-level speaker features while
keeping the simple d-vector architecture unchanged~\cite{li2017deep}. In the second direction, researchers pursue
end-to-end solutions which produce utterance-level speaker vectors directly~\cite{heigold2016end,zhang2016end,snyder2016deep,snyder2018xvector}.
A representative work in this direction is the x-vector architecture proposed by Snyder et al.~\cite{snyder2018xvector},
which produces the utterance-level speaker vectors (x-vectors) from the first- and second-order statistics
of the frame-level features.

For both the d-vector and x-vector architectures, however, there are two potential problems.
Firstly, the DNN models involve a parametric classifier (i.e., the last affine layer) during model training.
This means that part of the knowledge involved in the training data is used to learn a classifier
that will be ultimately thrown away during inference, leading to potential `information leak'.
Secondly, these models do not regulate the distribution of the derived speaker vectors, either at the frame-level
or at the utterance-level. The uncontrolled distribution will degrade the subsequent scoring component, especially
the PLDA model that assumes the speaker vectors are Gaussian~\cite{Ioffe06}.


\begin{figure*}[!htb]
    \centering
    \includegraphics[width=0.82\linewidth]{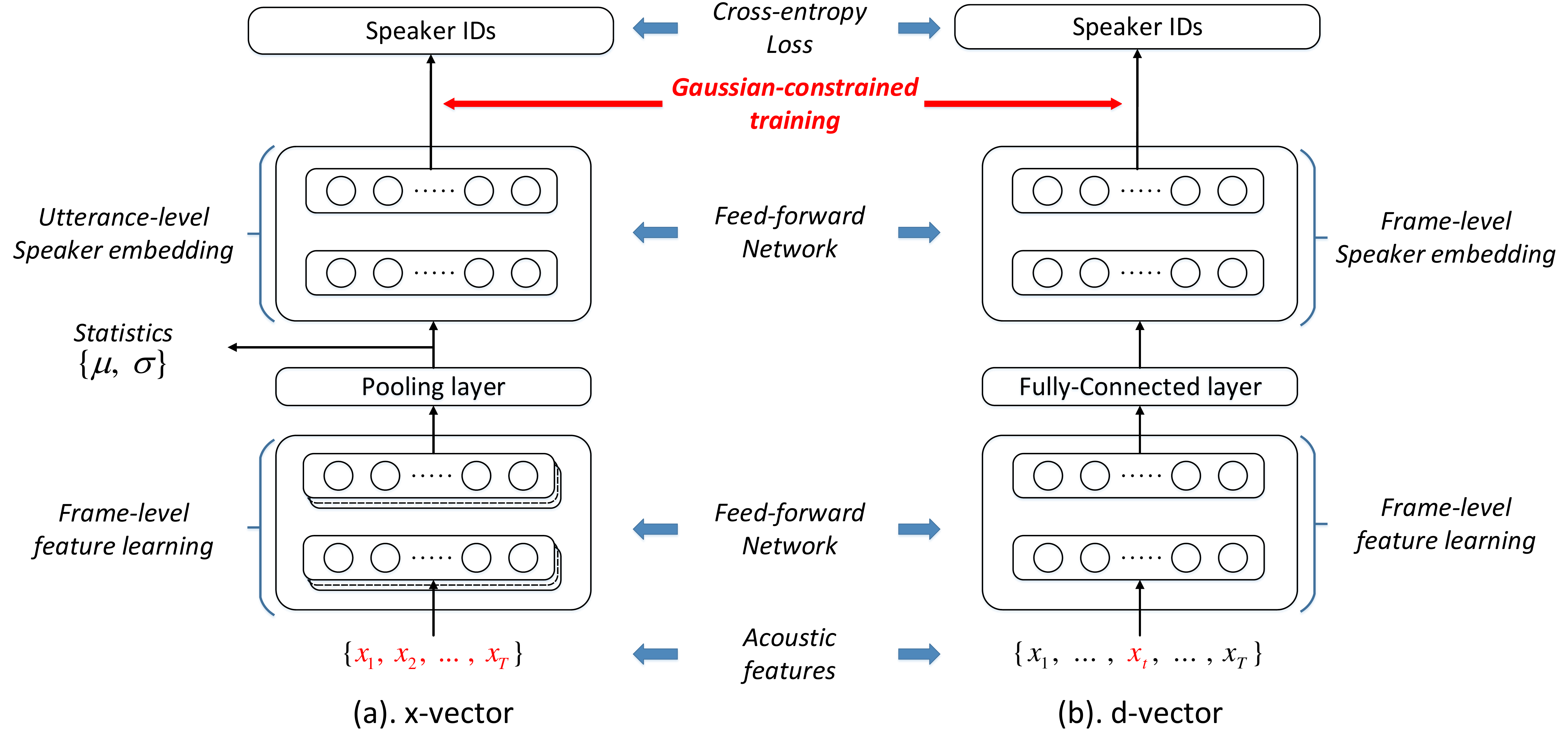}
    \vspace{-1mm}
    \caption{The architecture of the x-vector and d-vector models, shown in a comparative way.}
    \vspace{-2mm}
    \label{fig:model}
\end{figure*}

To deal with these two problems, we propose a Gaussian-constrained training approach in this paper. This new
training approach will (1) discard the parametric classifier to mitigate information leak, and (2)
enforce the distribution of the derived speaker vectors to be Gaussian to meet the requirement of the scoring component.
Our experiments on two databases demonstrated that the approach can
produce more representative and regular speaker vectors than both the d-vector and x-vector models,
which in turn leads to consistent performance improvement.

The organization of this paper is as follows.
Section~\ref{sec:deep} gives a brief overview for the x-vector and d-vector models, and Section~\ref{sec:method}
presents the proposed Gaussian-constrained training.
Experiments are reported in Section~\ref{sec:exp}, and the paper is concluded in Section~\ref{sec:cond}.

\section{Overview of x-vector and d-vector models}
\label{sec:deep}

The x-vector model and the d-vector model are two typical neural models adopted by ASV researchers.
The architectures of these models are shown in Figure~\ref{fig:model}, in a comparative way.

For the x-vector model, it consists of three components.
The first component is used for frame-level feature learning.
The input of this component is a sequence of acoustic features, e.g., filterbank coefficients.
After several feed-forward neural networks,
it outputs a sequence of frame-level speaker features.
The second component is a statistic pooling layer, in which statistics of the frame-level features, e.g., mean and standard deviation,
are computed. This statistic pooling maps a variable-length input to a fixed-dimensional vector.
The third component is used to produce utterance-level speaker vectors.
The output layer of this component is a softmax, in which each output node corresponds to a particular speaker in the training data.
The network is trained to discriminate all the speakers in the training data, conditioned on the input utterance.
Once the model is well trained, utterance-level speaker vectors, i.e., x-vectors, are read from a layer in the last component,
and a scoring model, such as PLDA, will be used to score the trials.

The d-vector model has a similar but simpler architecture. It consists of two components, one is for frame-level
feature learning, the same as the first component of the x-vector model, and the other is
for frame-level speaker embedding, the same as the third component of the x-vector model. Since the
entire architecture is frame-based, a pooling layer is not required.
The network is trained to discriminate speakers in the training data, but conditioned on each frame. Once the
training is done, utterance-level speaker vectors, i.e., d-vectors, are derived by averaging the frame-level features.
Finally, a scoring model such as PLDA will be used to score the trials.

\section{Gaussian-constrained training}
\label{sec:method}

As mentioned in Section~\ref{sec:intro}, both the d-vector and the x-vector models suffer from (1) information leak and (2) unconstrained distribution of
speaker vectors. We propose a Gaussian-constrained training approach to solve these problems.
In brief, this approach introduces a Gaussian prior on the output of the
embedding layer, which can be formulated as a regularization term in the training objective. Training with this augmented objective will enforce the
parameters of the classification layer more predictable, so more parameter-free. Meanwhile, it will urge the model producing more Gaussian speaker vectors,
at either the frame-level (for d-vector) or the utterance-level (for x-vector).

For a clear presentation, we shall use the x-vector model to describe the process, whereas the same argument applies to the
d-vector model in a straightforward way.
Specifically, if all the utterance-level x-vectors in the training set have been derived,
the speaker-level x-vector can be simply obtained by averaging all the utterance-level x-vectors
belong to that speaker. This is formally written as:

\begin{equation}
\label{eq:vs}
  v(s) = \frac{1}{|\mathcal{E}(s)|} \sum_{x \in \mathcal{E}(s)} f(x),
\end{equation}

\noindent where $\mathcal{E}(s)$ is the set of utterances belonging to speaker $s$;
$f(x)$ is the x-vector of utterance $x$; $v(s)$ is the speaker-level x-vector.

Based on the speaker-level x-vectors $\{v(s)\}$, each speech utterance $x$ can be easily classified as follows:

\begin{equation}
\label{eq:cos}
  p(s|f(x)) = \frac{e^{f(x) \cdot v(s)}}{\sum_{s'} e^{f(x) \cdot v(s')}},
\end{equation}

\noindent which can be regarded as a non-parametric classifier. If we use this non-parametric classifier to replace the
parametric classifier (usually the last affine layer) of the x-vector model in Fig.~\ref{fig:model}, we reach the
\emph{full-info training} proposed by Li et al.~\cite{li2018full}.

The model can be trained with the classical cross entropy (CE) criterion, written by:

\begin{equation}
\label{eq:cost}
  L' = - \sum_{t} \log \ p(s_t|f(x_t))
\end{equation}

\noindent where $x_t$ and $s_t$ are the $t$-th speech utterance and the corresponding ground truth label.
Note that the gradients on CE will be \emph{fully} propagated to the weights of the feature learning component, as the
parameters of the classifier (\ref{eq:cos}) are $\{v(s)\}$, which are dependent on the feature learning
component as well. This means all the knowledge of the data will be exploited to learn the feature component,
hence amending the information leak problem.

However, we cannot use a `virtual classifier' parameterized by $\{v(s)\}$ in a practical implementation; instead, we resort to
an engineering solution that designs a true parametric classifier, and replaces the parameters by $\{v(s)\}$ regularly.
This train-and-replacement scheme works well in many scenarios~\cite{li2018full}, but may slow the training or
cause fluctuation.

A more elegant approach is to keep the classifier parameters, but introduce a regularization term
that encourages the parameters approaching to $\{v(s)\}$. For this purpose, the following regularization term is designed:

\begin{equation}
\label{eq:cost}
  R = \sum_s \sum_{x_t \in \mathcal{E}(s) } ||f(x_t) - \theta_s||_2,
\end{equation}

\noindent where $\theta_s$ represent the parameters in the classifier that are associated with the
output node corresponding to speaker $s$.
With this regularization, the training objective is given by:

\[
L = L' + \alpha R,
\]
\noindent where $\alpha$ controls the strength of the regularization.
Clearly, if $\alpha$ is sufficiently large, $\theta_s$ will converge to the speaker-level x-vector $\{v(s)\}$.
Moreover, it is clear that the regularization term $R$ encourages all the utterance-level x-vectors $f(x_t)$
belonging to speaker $s$ to be a Gaussian  $N(\theta_s, \mathbf{I})$. Therefore, we name this new training approach
as \emph{Gaussian-constrained training}.

The Gaussian-constrained training possesses several advantages. Firstly,
it encourages the parameters of the classifier to converge to the speaker-level x-vectors, which equals to
removing these parameters gradually. This mitigates the information leak problem, but does not suffer from the unstability of
the full-info training. For this reason, Gaussian-constrained training can be regarded as a \emph{soft} full-info training.
Secondly, it encourages the utterance-level
x-vectors to be Gaussian, which is a key requirement for many scoring models, particularly PLDA.
Thirdly, this approach is flexible. We can choose $\alpha$ to control the strength of the regularization, or
choose other regularization forms to produce speaker vectors in other forms of distributions.


\vspace{-1mm}
\section{Experiments}
\label{sec:exp}
\vspace{-2mm}
\subsection{Data}
\vspace{-1mm}

Three datasets were used in our experiments: VoxCeleb, SITW and CSLT-SITW.
VoxCeleb was used for model training, while the other two were used for evaluation. More
information about these three datasets is presented below.

\emph{VoxCeleb}: A large-scale free speaker database collected by University of Oxford, UK~\cite{nagrani2017voxceleb}.
The entire database involves two parts: \emph{VoxCeleb1} and \emph{VoxCeleb2}.
Since part of the speakers is shared by VoxCeleb and SITW, a simple data purging was
conducted to remove all the data of the shared speakers. The purged dataset involves $1,236,567$ utterances from
$7,185$ speakers. This dataset was used to train both the d-vector model and the x-vector model, plus the LDA and PLDA models.
Data augmentation was applied, where the MUSAN corpus~\cite{musan2015} was used to generate noisy utterances and
the room impulse responses (RIRS) corpus~\cite{ko2017study} was used to generate reverberant utterances.

\emph{SITW}: A standard database used to test ASV performance in real-world conditions~\cite{mclaren2016speakers}.
It was collected from open-source media channels, and consists of speech data covering $299$ well-known persons.
There are two standard datasets for testing: \emph{Dev. Core} and \emph{Eval. Core}. We used these two sets to conduct the
first evaluation in our experiment.
Note that the acoustic condition of SITW is similar to that of the training set VoxCeleb, so this evaluation can
be regarded as a condition-matched evaluation.

\emph{CSLT-SITW}: A small dataset collected by CSLT at Tsinghua University. It consists of $11$ speakers,
each of which records a couple of Chinese digital strings by several mobiles. Each string contains $8$ Chinese digits,
and the duration is about $2$-$3$ seconds. The scenarios involve laboratory, corridor, street, restaurant, bus and subway.
Speakers varied their poses during the recording, and the mobile phones were placed both near and far.
There are $6,915$ utterances in total.

\subsection{Settings}

For a comprehensive comparison,
three baseline systems following the Kaldi SITW recipe~\cite{povey2011kaldi} were built:
an i-vector system, an x-vector system and a d-vector system.

For the i-vector system, the acoustic feature involves $24$-dimensional MFCCs plus the log energy,
augmented by the first- and second-order derivatives. We also apply cepstral mean normalization (CMN)
and the energy-based voice active detection (VAD).
The UBM consists of $2,048$ Gaussian components,
and the dimensionality of the i-vector space is $400$.
LDA is applied to reduce the dimensionality of i-vectors to $150$ prior to PLDA scoring.

For the x-vector system, the feature-learning component is a 5-layer time-delay neural network (TDNN). The slicing parameters for the
five TD layers are:
\{$t$-$2$, $t$-$1$, $t$, $t$+$1$, $t$+$2$\}, \{$t$-$2$, $t$, $t$+$2$\}, \{$t$-$3$, $t$, $t$+$3$\}, \{$t$\}, \{$t$\}.
The statistic pooling layer computes the mean and standard deviation of the frame-level features from a speech segment.
The size of the output layer is $7,185$, corresponding to the number of speakers in the training set.
Once trained, the $512$-dimensional activations of the penultimate hidden layer are read out as an x-vector.
This vector is then reduced to a $150$-dimensional vector by LDA, and finally the PLDA model is employed to score the trials.
Refer to~\cite{snyder2017deep} for more details.
In the Gaussian-constrained training, the hyper-parameter $\alpha$ is empirically set to $0.05$.

For the d-vector system, the DNN structure is similar to that of the x-vector system.
The only difference is that the statistic pooling layer in the x-vector model is replaced
by a TD layer whose slicing parameter is set to \{$t$-$3$, $t$, $t$+$3$\}.
Once trained, the $512$-dimensional deep speaker features are derived from the output of the penultimate hidden layer,
and the utterance-level d-vectors are obtained by average pooling.
Similarly, the d-vectors are transformed to $150$-dimensional vectors by LDA, and the PLDA model is employed to score the trials.
The hyper-parameter of the Gaussian-constrained training is empirically set to $0.01$.

\vspace{-2mm}
\subsection{Results}

\subsubsection{SITW}
\vspace{-1mm}

The results on the two SITW evaluation sets, \emph{Dev. Core} and \emph{Eval. Core}, are reported
in Table~\ref{tab:dev} and Table~\ref{tab:eval}, respectively. The results are reported in terms of three metrics:
the equal error rate (EER), and the minimum of the normalized detection cost function (minDCF) with two settings:
one with the prior target probability $P_{tar}$ set to $0.01$ (DCF($10^{-2}$)), and the other with $P_{tar}$
set to $0.001$ (DCF($10^{-3}$)).

From these results, it can be observed that the proposed Gaussian-constrained training improves
both the x-vector and d-vector systems, in terms of all these three metrics.
Furthermore, this approach seems more effective for the x-vector system.
A possible reason is that for the d-vector system, the average pooling may corrupt the
Gaussian property. Another possible reason is that
the frame-level constraint in the d-vector system may lead to unstable parameter update compared to
the utterance-level constraint in the x-vector system. Nevertheless, more investigation is required to
understand this discrepancy.

\vspace{-4mm}

\begin{table}[htb!]
 \begin{center}
  \caption{Performance on SITW Dev. Core.}
   \label{tab:dev}
     \begin{tabular}{|l|c|c|c|}
       \hline
             Embedding         &  DCF($10^{-2}$)    &   DCF($10^{-3}$)  &   EER(\%) \\
       \hline
             i-vector          &  0.4279            &   0.5734          &   4.967   \\
       \hline
             d-vector          &  0.4875            &   0.6837          &   5.314   \\
       d-vector + \emph{Gauss} &  0.4861            &   0.6812          &   5.160  \\
       \hline
             x-vector          &  0.3025            &   0.4862          &   2.965   \\
       x-vector + \emph{Gauss} & \textbf{0.2826}    & \textbf{0.4551}   & \textbf{2.734}   \\
       \hline
     \end{tabular}
 \end{center}
\end{table}

\vspace{-11mm}

\begin{table}[htb!]
 \begin{center}
  \caption{Performance on SITW Eval. Core.}
   \label{tab:eval}
     \begin{tabular}{|l|c|c|c|}
      \hline
             Embedding         &  DCF($10^{-2}$)    &   DCF($10^{-3}$)  &   EER(\%) \\
       \hline
             i-vector          &  0.4577            &   0.6214          &   5.249   \\
       \hline
             d-vector          &  0.5206            &   0.7570          &   5.686   \\
       d-vector + \emph{Gauss} &  0.5149            &   0.7496          &   5.659  \\
       \hline
             x-vector          &  0.3235            &   0.4875          &   3.390   \\
       x-vector + \emph{Gauss} &  \textbf{0.3032}   &  \textbf{0.4520}  &  \textbf{3.034}  \\
       \hline
     \end{tabular}
 \end{center}
\end{table}

\vspace{-8mm}

\subsubsection{CSLT-SITW}
\vspace{-1mm}
The performance on the CSLT-SITW set is reported in Table~\ref{tab:cslt}. Note that the acoustic properties and linguistic conditions are
clearly different from the training data. From Table~\ref{tab:cslt}, it can be observed that in spite of this mismatch,
the Gaussian-constrained training still delivers consistent performance improvement on both the x-vector and d-vector systems, at least
in terms of EER and DCF($10^{-2}$). The strange degradation in DCF($10^{-3}$) may be attributed to the fact that this
new training approach emphasizes on a different operation point, though more analysis is required.

\vspace{-4mm}

\begin{table}[htb!]
 \begin{center}
  \caption{Performance on CSLT-SITW.}
   \label{tab:cslt}
     \begin{tabular}{|l|c|c|c|}
      \hline
             Embedding         &  DCF($10^{-2}$)    &   DCF($10^{-3}$)  &   EER(\%) \\
       \hline
             i-vector          &  0.4425            &   0.5698          &   6.479   \\
       \hline
             d-vector          &  0.3881            &   0.4584          &   5.494   \\
       d-vector + \emph{Gauss} &  0.3706            &   0.4701          &   5.297   \\
       \hline
             x-vector          &  0.2731            &  \textbf{0.3227}  &   4.139   \\
       x-vector + \emph{Gauss} &  \textbf{0.2418}   &  0.3746           &  \textbf{3.474}   \\
       \hline
     \end{tabular}
 \end{center}
\end{table}

\vspace{-6mm}

\section{Conclusions}
\label{sec:cond}

This paper proposed a Gaussian-constrained training that can be applied to both the feature-based
system (d-vector) and utterance-based system (x-vector) for ASV. The basic idea is to enforce a parameter-free classifier
so that all the knowledge of the training data would be learned by the feature component; additionally, it encourages
the derived speaker features, at either frame-level or utterance-level, to be Gaussian. The former allows more
effective usage of the training data, and the latter boosts the PLDA scoring. The experimental results
demonstrated that the proposed approach can deliver consistent performance improvement, not only on matched
data, but also on non-matching conditions. As for the future work, more comprehensive analysis will be conducted to understand
the behavior of the Gaussian-constrained training, e.g., the impact of the constraint on different layers.



\bibliographystyle{IEEEbib}
\bibliography{refs}

\end{document}